\documentclass[aps,pra,showpacs,twocolumn,groupedaddress,a4paper,amsmath]{revtex4}

\usepackage{graphics}

\begin{document}

%\preprint{}

\title{The axial breathing mode in rapidly rotating Bose-Einstein condensates and uncertainty of the rotation velocity}


\author{Gentaro Watanabe$^{a,b,c}$}

%\thanks{}
\affiliation{$^{a}$NORDITA, Blegdamsvej 17, DK-2100 Copenhagen \O, Denmark
\\
$^{b}$CNR-INFM BEC, Department of Physics, University of Trento, Via Sommarive 14, 38050 Povo (TN) Italy
\\
$^{c}$The Institute of Chemical and Physical Research (RIKEN), 2-1 Hirosawa, Wako, Saitama 351-0198, Japan}


%\date{\today}

\begin{abstract}
Experiments on the axial breathing mode in a rapidly rotating
Bose-Einstein condensate are examined.  Assuming a cold cloud without
thermal component, we show that errors due to defocus of an imaging
camera in addition to an inclination of the rotational axis can lead
to a significant underestimate of the rotation rate in the fast
rotation limit; within these uncertainties, our theoretical prediction
agrees with the experimental data.  We also show that, in the fast
rotation regime, the Thomas-Fermi theory, which is inapplicable there,
underestimates the rotation rate.  Underestimate of the rotation rate
due to these effects would also partly explain a discrepancy between
theory and experiment for the Tkachenko mode frequency in the fast
rotation regime.
\end{abstract}

%\keywords{}

\maketitle


The creation of vortices in atomic Bose-Einstein condensates has
opened up a new horizon in the study of superfluids.
This system enables us to investigate an unexplored regime,
where the vortex core is comparable to the intervortex distance
since the interaction energy per particle $\sim gn$ is much smaller
than that of liquid He-II 
(see, e.g., Ref.\ \cite{coddington} and references therein).  
Here $n$ is the particle number density, 
$g\equiv 4\pi\hbar^2 a_{\rm s}/m$ is the two-body interaction strength,
$m$ is the particle mass, and $a_{\rm s}$ is the $s$-wave scattering length.

For a harmonically trapped Bose-Einstein condensate, 
when the rotation angular velocity $\Omega$ is close to
the transverse trap frequency $\omega_{\perp}$,
so that $\hbar\Omega \agt gn$, the condensate wave function is dominated by 
the lowest Landau level (LLL) component \cite{ho}.
When the number of particles $N$ is much larger than 
the number of vortices $N_{\rm v}$, the system may be described by
the Gross-Pitaevskii equation \cite{cooper,sinova}.
Recently, Schweikhard {\it et al.} \cite{schweikhard} have reached what we
shall refer to as the mean-field LLL regime 
\cite{ho,wbp,ckr,baym_tk,aftalion,sonin,cst,wsgp,elastic}, 
in which $\hbar\Omega \agt gn$ with $N\gg N_{\rm v}$.

In Ref.\ \cite{schweikhard}, Schweikhard {\it et al.} have measured
the axial breathing mode frequency and they have observed a frequency shift
in the rapid rotation regime,
which cannot be described by the superfluid hydrodynamic models of 
Refs.\ \cite{sedrakian,cozzini}. 
We have studied breathing modes in the mean-field LLL regime
in Ref.\ \cite{breathing}, and have pointed out that our theoretical framework
developed there could describe this frequency shift.
The purpose of this paper is to compare in detail 
our theoretical prediction
of the axial breathing mode frequency with experimental data.

In the Appendix of Ref.\ \cite{breathing}, in which we discussed 
the breathing modes in three dimensions,
we employed a generalized LLL wave function $\phi_{\rm ex}$,
in which the transverse oscillator length is treated as a variable.
There we assumed the coarse-grained density profile 
$n({\bf r})=N\langle |\phi_{\rm ex}|^2 \rangle$ to be 
a Thomas-Fermi (TF) parabola
in the transverse direction and a Gaussian form
in the axial ($z$) direction, 
which is correct
in the fast rotation limit, in which the cloud is so dilute that 
$\hbar\omega_z\gg gn$, where $\omega_z$ is the axial trap frequency.
In the present work, we employ an extended version of this framework, 
which describes the coarse-grained density profile of the cloud more correctly
at lower rotation velocities.
Here we assume 
\begin{equation}
  n({\bf r})
  = n(0) \left(1-\frac{r_{\perp}^2}{R_{\perp}^2}\right)
  \left(1-\frac{z^2}{\alpha R_z^2}\right)^\alpha\ ,\label{gtf}
\end{equation}
where $r_{\perp}^2=x^2+y^2$, 
$R_{\perp}$ is the TF radius in the transverse direction,
$\alpha$ is a parameter (static), which determines the shape of the axial
density profile, and 
$\sqrt{\alpha}R_z$ is the cloud radius in the axial direction.
To satisfy the normalization condition, $\int d^3r\ n({\bf r})=N$, we set 
$n(0)=2N/[\pi\sqrt{\alpha}B\left(\alpha+1,\frac{1}{2}\right)R_{\perp}^2R_z]$,
where $B$ is the beta-function defined as 
$B(p,q)\equiv \int_0^1 dt\ t^{p-1}(1-t)^{q-1}$.
The axial density profile of Eq.~ (\ref{gtf}) interpolates between
the one-dimensional TF parabola $(\alpha=1)$ and the Gaussian form 
$(\alpha=\infty)$.

The extended LLL wave function corresponding to the density profile
(\ref{gtf}) is
$\phi_{\rm ex}({\bf r}) = A_{\rm ex} \prod_{i=1}^{N_{\rm v}}(\zeta-\zeta_i)\
\exp{\left[-\left(\lambda^{-2} -i\beta\right)\, 
r_{\perp}^2/2d_{\perp}^2\right]}\ 
\bigl[1-(z^2/\alpha R_z^2)\bigr]^{\alpha/2}
\exp{\bigl[i\gamma z^2/2d_z^2\bigr]}$, 
where $\zeta=x+iy$, $\zeta_i$ are the vortex positions, 
$A_{\rm ex}$ is the normalization constant, 
$d_{\perp}\equiv\sqrt{\hbar/m\omega_{\perp}}$ and 
$d_z\equiv\sqrt{\hbar/m\omega_z}$
are the transverse and the axial 
oscillator lengths, respectively.
Here 
the dynamical variable $\lambda$ describes 
the variation of the effective transverse oscillator length.
The dynamical variables $\beta$ and $\gamma$ 
generate a radial and an axial velocity field, respectively, 
which cause the homologous change of the density profile.

Following the same way as in Ref.\ \cite{breathing}, we obtain 
the Lagrangian function for the above $\phi_{\rm ex}$ as
\begin{align}
  {\cal L}[\phi_{\rm ex}]=& -\frac{\hbar}{2}\left(\frac{\dot{\beta}}{3}X^2
+\frac{\dot{\gamma}}{2} h(\alpha)Z^2\right)\nonumber\\
&-\biggl[\frac{\hbar\omega_{\perp}}{2}
\biggl\{\frac{3l^2}{X^2}
+\left(\beta^2+1\right)\frac{X^2}{3}\biggr\}
+\frac{\hbar\omega_z}{2}\biggl\{\frac{\mbox{\sl g}(\alpha)}{2Z^2}\nonumber\\
&+\left(\gamma^2+1\right) h(\alpha)\frac{Z^2}{2}\biggr\}
+\frac{4\hbar\omega_{\perp}}{3} f(\alpha)\frac{\kappa}{X^2 Z}\biggr]\ ,\label{lagrangian}
\end{align}
(the dot denotes the time derivative) with 
$f(\alpha)\equiv \sqrt{2\pi/\alpha} B\bigl(2\alpha+1, \frac{1}{2}\bigr)/B^2\bigl(\alpha+1, \frac{1}{2}\bigr)$, 
$\mbox{\sl g}(\alpha)\equiv 2\alpha B\bigl(\alpha-1, \frac{3}{2}\bigr)/B\bigl(\alpha+1, \frac{1}{2}\bigr)$, 
$h(\alpha)\equiv 2\alpha B\bigl(\alpha+1, \frac{3}{2}\bigr)/B\bigl(\alpha+1, \frac{1}{2}\bigr)$, 
and the dimensionless interaction strength
$\kappa \equiv (\sqrt{2\pi}d_z)^{-1} (mbgN)/(2\pi\hbar^2)$, 
where 
$b\equiv\langle |\phi_{\rm ex}|^4 \rangle/\langle |\phi_{\rm ex}|^2 \rangle^2$ 
is the Abrikosov parameter, which is comparable to unity.
Here $X\equiv R_{\perp}/d_{\perp}$, $Z\equiv R_z/d_z$, 
and $l$ is defined by the expectation value
$l_z$ of the angular momentum in the $z$ direction per particle as
$l\equiv (l_z/\hbar)+1=X^2/3\lambda^2$.
Note that, in the limit of the axial density profile being Gaussian 
($\alpha\rightarrow\infty$), the functions 
$f(\alpha),\ \mbox{\sl g}(\alpha)$, and $h(\alpha)$ are unity
and thus Eq.\ (\ref{lagrangian}) reduces to Eq.\ (A8) 
in Ref.\ \cite{breathing}.

Using the variational Lagrangian formalism, we obtain 
coupled equations similar to Eqs.\ (A.12) and (A.13) in Ref.\ \cite{breathing} 
that determine $R_{\perp}$ and $R_z$ (or $\alpha$) in the equilibrium state: 
\begin{align}
  &\left(1-\frac{\Omega_0^2}{\omega_{\perp}^2}\right)X_0^4
-8\kappa f(\alpha)Z_0^{-1}=0\ ,\label{eq1}\\
  &h(\alpha)Z_0^3-\mbox{\sl g}(\alpha)Z_0^{-1}
-\frac{8\kappa}{3}f(\alpha)\frac{\omega_{\perp}}{\omega_z}X_0^{-2}=0\ ,
\label{eq2}
\end{align}
where $X_0$ and $Z_0$ are the equilibrium values of
$X$ and $Z$, and $\Omega_0$ is the rotation velocity in the equilibrium state,
which we obtain as $\Omega_0/\omega_{\perp}=3l/X_0^2$.
Eliminating $\kappa$ from Eqs.\ (\ref{eq1}) and (\ref{eq2}), 
we write $\Omega_0$ as
\begin{equation}
  \frac{\Omega_0}{\omega_\perp}=\sqrt{1-3\frac{\omega_z}{\omega_{\perp}}
\frac{1}{X_0^2 Z_0^2}\left(h(\alpha)Z_0^4-\mbox{\sl g}(\alpha)\right)}\ .
\label{rotvel}
\end{equation}

In addition to Eqs.\ (\ref{eq1}) and (\ref{eq2}), we get an extra equation
by minimizing the energy with respect to $\alpha$:
\begin{equation}
  h'(\alpha) Z_0^3 + \mbox{\sl g}'(\alpha) Z_0^{-1} + \frac{16\kappa}{3}f'(\alpha)\frac{\omega_{\perp}}{\omega_z}X_0^{-2}=0\ ,
\label{eq3}
\end{equation}
where the prime denotes the derivative with respect to $\alpha$.
Using Eqs.\ (\ref{eq2}) and (\ref{eq3}), $Z_0$ is written
as a function of only $\alpha$:
\begin{equation}
  Z_0=\left(2\frac{\mbox{\sl g}(\alpha)}{f(\alpha)}-\frac{1}{2}\frac{\mbox{\sl g}'(\alpha)}{f'(\alpha)}\right)^{1/4}\left(2\frac{h(\alpha)}{f(\alpha)}+\frac{1}{2}\frac{h'(\alpha)}{f'(\alpha)}\right)^{-1/4} .
\label{z0}
\end{equation}

Resulting expression of the mode frequency is
\begin{widetext}
\begin{equation}
  \omega^2 
  = \frac{1}{2}
  \left[\left(3+\frac{\mbox{\sl g}(\alpha)}{h(\alpha)}\frac{1}{Z_0^4}\right)\omega_z^2+4\omega_{\perp}^2\right]
  \pm\frac{1}{2}\sqrt{
  \left[\left(3+\frac{\mbox{\sl g}(\alpha)}{h(\alpha)}\frac{1}{Z_0^4}\right)\omega_z^2-4\omega_{\perp}^2\right]^2
  + 8\omega_{\perp}^2\omega_z^2 \left(1-\frac{\mbox{\sl g}(\alpha)}{h(\alpha)}\frac{1}{Z_0^4}\right)\left(1-\frac{\Omega_0^2}{\omega_{\perp}^2}\right)
  }\ .
\label{omega3d}
\end{equation}
\end{widetext}
The $1/Z_0^4$ terms come from the zero-point energy
in the $z$ direction, which is not taken into account 
in the hydrodynamic calculations \cite{sedrakian,cozzini} 
based on the TF theory \cite{note slow}.
We see that, in the limit of $\alpha\rightarrow\infty$, Eq. (\ref{omega3d}) 
reproduces Eq.\ (A16) in Ref.\ \cite{breathing}, which
is obtained for a cloud with a Gaussian form of the axial density profile
\cite{note_error}.
Note that Eq.\ (\ref{omega3d}) can describe 
the frequency shift of the axial breathing mode from 
$\sqrt{3}\omega_z$ to $2\omega_z$ with increasing $\Omega_0$ 
in the fast rotation regime reported in Ref.\ \cite{schweikhard}.
For values of $\Omega_0$ at which the frequency shift occurs,
the second term in the square root of 
Eq.\ (\ref{omega3d}) can be neglected and this equation reduces to
\begin{equation}
\omega_{\rm tr}^2 \simeq 4\omega_{\perp}^2\quad \mbox{and}\quad
\omega_{\rm ax}^2 \simeq\left(3+\displaystyle{\frac{\mbox{\sl g}(\alpha)}{h(\alpha)}\frac{1}{Z_0^4}}\right)\omega_z^2\ ,
\label{omega3d_rapid}
\end{equation}
where $\omega_{\rm tr}$ and $\omega_{\rm ax}$ are the transverse and axial 
breathing mode frequency, respectively.
Since $Z=R_z/d_z\ge 1$,
the first case of Eq.\ (\ref{omega3d_rapid}) is given by the upper sign 
of Eq.\ (\ref{omega3d}) 
and the second one by the negative sign for $\omega_{\perp}>\omega_z$ 
as in the experiment of Ref.\ \cite{schweikhard}.
If the system is in the LLL limit, $z$ dependence of the wave function
corresponds to the ground state of a particle in a harmonic potential, 
which is reproduced by $\alpha\rightarrow\infty$ giving 
$n({\bf r})\propto \exp{\left(-z^2/R_z^2\right)}$ with $R_z=d_z$, then 
$Z_0\simeq1$, and $\mbox{\sl g}(\infty)=h(\infty)=1$.
Thus the second expression of Eq.\ (\ref{omega3d_rapid}) 
leads to $\omega_{\rm ax}\simeq 2\omega_z$.  Otherwise $1/Z_0^4$ term is small
and this expression gives $\omega_{\rm ax}\simeq \sqrt{3}\omega_z$.
This observation shows that the zero-point energy in the $z$ direction,
which is neglected in the TF theory,
plays an essential role in the frequency shift of the axial breathing mode.
We also note that, from the first case of Eq.\ (\ref{omega3d_rapid}),
such a change in the mode frequency does not exist for 
the transverse breathing mode \cite{note transverse}.

In comparing experimental data and theoretical prediction, a key issue 
is the determination of the rotation velocity $\Omega_0$.
In Ref.\ \cite{schweikhard}, $\Omega_0$
is evaluated from the aspect ratio of the cloud using the TF 
theory, which is invalid in the fast rotation regime.
Thus we do not use the data of $\Omega_0$ given in Ref.\ \cite{schweikhard};
instead, we determine $\Omega_0$ using the present theoretical framework.
In the experiment of Ref.\ \cite{schweikhard}, what Schweikhard {\it et al.} 
have observed is the phase contrast image of the cloud, 
which is proportional to the optical density 
$n_{\rm opt}(y,z)=\int dx~ n({\bf r})$ \cite{data}. 
To perform direct comparison with the present theory, 
Schweikhard reanalyzed the stored data of the image obtained in 
that experiment and
newly measured the rms radii of the optical density profile \cite{data}.

The rms radii are extracted in the following way.
Since the signal to noise ratio is low in the fast rotation regime 
(see, e.g., Fig.\ 1(c) in Ref.\ \cite{schweikhard}),
contribution from the outer region of the cloud will produce a large error
of the rms radii. To make the measured values immune against the noise in the 
images, image pixels with an optical density lower than 
50 \% of the peak optical density had to be discarded \cite{data}.
Thus the rms radii $\langle y^2 \rangle^{1/2}$ and 
$\langle z^2 \rangle^{1/2}$ are measured for a part of the cloud in which
the optical density is higher than 50 \% of its peak value \cite{data},
i.e.,
\begin{equation}
  \langle y^2 \rangle = \frac{\int_{\cal R}dy dz\ y^2 n_{\rm opt}(y,z)}{\int_{\cal R}dy dz\ n_{\rm opt}(y,z)}\ ,
\label{rmsrad}
\end{equation}
where the integration region 
${\cal R}=\{(y,z)\ |\ n_{\rm opt}(y,z)\ge 0.5 n_{\rm opt}(0,0)\}$
($\langle z^2 \rangle$ is given by an analogous expression).

On the other hand, a theoretical expression of the optical density 
for the density profile of Eq.\ (\ref{gtf}) is
\begin{equation}
  n_{\rm opt}(y,z)=\frac{4}{3} n(0)R_{\perp}
\left(1-\frac{y^2}{R_{\perp}^2}\right)^{3/2} \left(1-\frac{z^2}{\alpha R_z^2}\right)^\alpha\ .
\label{noptgtf}
\end{equation}
Similar to the above experimental procedure, theoretical values of the 
rms radii are calculated from Eq.\ (\ref{rmsrad}) with Eq.\ (\ref{noptgtf}).
We note that since the equilibrium values of $R_z$ and $\alpha$ 
are related by Eq.\ (\ref{z0}), $\langle y^2 \rangle$ and $\langle z^2 \rangle$
given by the present theory are functions of $R_{\perp}$ and 
$\alpha$ (or $R_z$).

Writing the experimental data of $\langle y^2 \rangle$ and 
$\langle z^2 \rangle$ as $\langle y^2 \rangle_{\rm ex}$ and 
$\langle z^2 \rangle_{\rm ex}$,
and those of the theoretical prediction as 
$\langle y^2 \rangle_{\rm th}$ and 
$\langle z^2 \rangle_{\rm th}$, for each shot of the experiment
$R_{\perp}$ and $\alpha$ are determined by minimizing
$\Delta I_{y^2}(R_{\perp},\alpha)\equiv \left(\langle y^2 \rangle_{\rm th}-\langle y^2 \rangle_{\rm ex}\right)^2$ and 
$\Delta I_{z^2}(R_{\perp},\alpha)\equiv \left(\langle z^2 \rangle_{\rm th}-\langle z^2 \rangle_{\rm ex}\right)^2$ simultaneously.
From the obtained value of $\alpha$, $R_z$ (i.e., $Z_0$) is readily calculated 
by Eq.\ (\ref{z0}), and then $\Omega_0$ is determined by Eq.\ (\ref{rotvel}) 
using these values of $R_{\perp}, R_z$, and $\alpha$.
Theoretical value of the mode frequency is calculated 
from Eq.\ (\ref{omega3d}).
We use the same values of the experimental
parameters as those in Ref.\ \cite{schweikhard}, i.e.,
$\omega_{\perp}=2\pi\times 8.3$ Hz ($d_{\perp}\simeq3.7$ $\mu$m), 
$\omega_z=2\pi\times 5.3$ Hz ($d_z\simeq4.7$ $\mu$m), and
$a_{\rm s}\simeq5.6$ nm for the triplet state of $^{87}$Rb.

\begin{figure}[t]
%\begin{center}\vspace{0.0cm}
\rotatebox{0}{
\resizebox{8.2cm}{!}
{\includegraphics{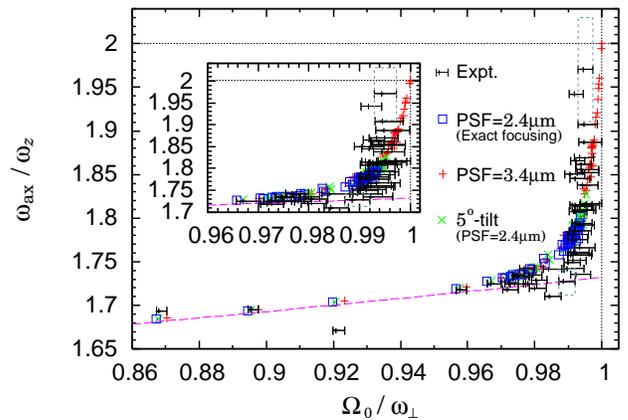}}}
\caption{\label{fig_omega}(Color online)
  Axial breathing mode frequency $\omega_{\rm ax}$ as a function of $\Omega_0$.
  The horizontal error bars are the experimental data and the points show 
  our theoretical results (see text for details).
  The dashed line shows the result of 
  hydrodynamic models by Refs.\ \cite{sedrakian,cozzini}.
  }
%\end{center}
\end{figure}

There are two major sources of error in the experimental data of the rms radii 
\cite{data}.
One is an overestimate of $\langle y^2\rangle_{\rm ex}^{1/2}$ and 
$\langle z^2\rangle_{\rm ex}^{1/2}$ due to defocusing of the imaging camera.
In the experiment, it cannot be completely excluded to have the camera 
defocused by one depth of focus, 
i.e., the imaging resolution gets worse by a factor of $\sqrt{2}$ 
at most \cite{data}.
When the focusing is exact, an effect of the finite imaging resolution 
is corrected by subtracting in quadrature the rms width of the Gaussian 
point spread function (PSF) of $2.4$ $\mu$m from the raw data of 
the rms radii.
Defocusing by one depth of focus corresponds to a point spread function of
$3.4$ $\mu$m instead of $2.4$ $\mu$m \cite{data}.
Since $\langle z^2\rangle_{\rm ex} \ll \langle y^2\rangle_{\rm ex}$
at $\Omega_0$ on which we are focusing,
the ratio of increase of $\langle z^2\rangle_{\rm ex}^{1/2}$ 
due to the defocusing
is larger than that of $\langle y^2\rangle_{\rm ex}^{1/2}$.
Thus this effect makes an observed image more prolate than reality
and leads to an underestimate of $\Omega_0$.

The other source is an overestimate of $\langle z^2\rangle_{\rm ex}^{1/2}$ 
due to a tilt of the rotation axis towards/away from the imaging camera.
The tilt angle in the direction of the camera cannot be measured and
thus the uncertainty of the tilt angle produces an error of 
$\langle z^2\rangle_{\rm ex}^{1/2}$. 
Judging from the observed inclination in the plane perpendicular to the camera,
the tilt towards/away from the camera is estimated at up to $\sim 5^\circ$ 
\cite{data}. 
This effect also leads to an underestimate of $\Omega_0$ due to the same reason
as mentioned above.

In Fig.\ \ref{fig_omega}, we plot the resulting axial breathing mode frequency
$\omega_{\rm ax}$ as a function of $\Omega_0$.
The horizontal error bars show the experimental results:
the left end of each error bar corresponds to the case of the exact focusing 
given by the PSF of $2.4$ $\mu$m, the right end to the case with 
the PSF of 3.4 $\mu$m, and the short vertical bar in the middle to
the case with the PSF of $2.4$ $\mu$m and the rotation axis 
being inclined towards/away from the camera by $5^\circ$ \cite{note_tilt}.
The two dotted boxes show the range of errors in $\omega_{\rm ax}$ and $\Omega_0$
for the two shots of the experiment.
Theoretical results are shown by points: 
the blue squares for the PSF of $2.4$ $\mu$m (exact focusing), 
the red ``$+$'' signs for the PSF of $3.4$ $\mu$m, and 
the green ``$\times$'' signs for the case with the axial tilt by $5^\circ$ and
the PSF of $2.4$ $\mu$m (here we referred to as the $5^\circ$-tilt case).
We observe that the theoretical results for all the three cases 
are almost on a single line irrespective of the way of correcting the data.

According to Fig.\ \ref{fig_omega}, the effect of defocusing is large enough 
that our theoretical prediction and experimental data agree 
within the error due to this effect.
The effect of the inclination of the rotation axial, on the other hand, 
is not so large that it can explain the discrepancy between 
theory and experiment by itself. 
However, in the fast rotation regime, where the frequency shift occurs,
this effect gives a significant correction in $\Omega_0$ of 
$\sim 0.002\omega_{\perp}$ 
because the cloud is very oblate with the aspect ratio of $\sim 0.2$.

\begin{figure}[t]
%\begin{center}\vspace{0.0cm}
\rotatebox{0}{
\resizebox{8.2cm}{!}
{\includegraphics{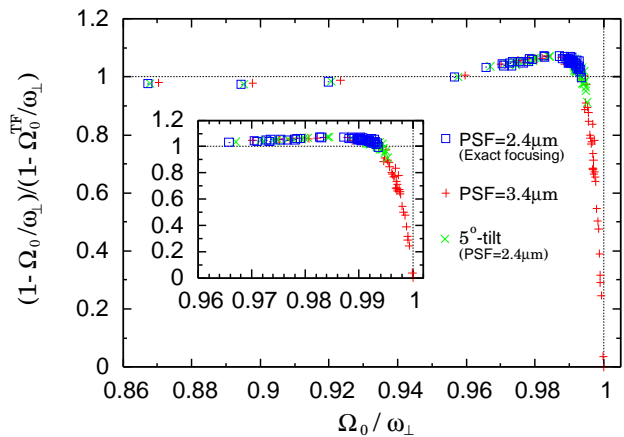}}}
\caption{\label{fig_omega0}(Color online)
  Comparison between the rotation velocity $\Omega_0$ 
  determined in the present analysis and $\Omega_0^{\rm TF}$
  obtained by the TF theory.
  }
%\end{center}
\end{figure}

It is instructive to compare the rotation velocity determined here, $\Omega_0$,
with that obtained by the TF theory, $\Omega_0^{\rm TF}$, 
which is invalid when the zero-point energy in the 
$z$ direction is important. 
From the same data of 
$\langle y^2\rangle_{\rm ex}$ and $\langle z^2\rangle_{\rm ex}$, 
we obtain $\Omega_0^{\rm TF}$ in the same way as $\Omega_0$ 
using the three-dimensional TF profile whose optical density is
$n_{\rm opt}(y,z)=(4/3) n(0) R_{\perp} 
(1-y^2/R_{\perp}^2-z^2/R_z^2)^{3/2}$, 
and a relation between the rotation rate and 
the aspect ratio given by the TF theory, 
$\Omega_0^{\rm TF}/\omega_{\perp}
=[1-(R_z/d_z)^2/(R_{\perp}/d_{\perp})^2]^{1/2}$,
instead of Eq.\ (\ref{rotvel}).
The absolute value of $\Omega_0^{\rm TF}$ itself is 
rather close to that of $\Omega_0$: difference between them is
at most $\alt 4$ \%.
However, the difference of this amount matters at fast rotation rates,
where $\omega_{\rm ax}$ shows the rapid change.
Thus we compare $1-\Omega_0/\omega_{\perp}$ and 
$1-\Omega_0^{\rm TF}/\omega_{\perp}$ in Fig.\ \ref{fig_omega0}. 
Note a striking drop at $0.995\alt \Omega_0/\omega_{\perp} \le 1$, 
which shows the TF theory underestimates the rotation velocity in this region.

Finally we should mention effects of thermal component, which we have
neglected in our analysis.  Due to the low signal to noise ratio in
the fast rotation regime, it is hard to measure the fraction of the
thermal component and we cannot completely exclude its existence \cite{data}. 
Thus it is possible that the thermal component whose eigenfrequency is 
$2\omega_z$ drags the oscillation of the condensate and could act to 
increase the oscillation frequency of the whole cloud 
towards $2\omega_z$ \cite{data}.

In this work, we have analyzed the axial breathing mode frequency 
using new data of the rms radii obtained in a follow-up measurement 
with attention to remove effects of noise \cite{data}. 
We have shown that the effects of defocus of the imaging camera
and an inclination of the rotation axis lead to a significant underestimate
of $\Omega_0$; 
especially the former one is so large that, within an uncertainty
due to this effect, our theoretical prediction agrees 
with the experimental data. 
We have also shown that TF theory tends to underestimate $\Omega_0$
in the fast rotation regime, where this theory is no longer valid.
Since these effects also exist in the experiment of the Tkachenko mode 
\cite{schweikhard},
they would partly explain a discrepancy between theory and experiment for 
the Tkachenko mode frequency in the fast rotation regime \cite{sonin,cst}.

%\vspace*{\fill}
%\begin{acknowledgements}
The author is very grateful to Volker Schweikhard for generously
providing us with his experimental data, for reanalyzing the data, 
and for many valuable comments.
He also thanks Chris Pethick for helpful discussion and comments.
This work was supported by JSPS.
%\end{acknowledgements}

%\newpage


\begin{thebibliography}{99}
\bibitem{note slow} With neglect of the $1/Z_0^4$ terms, Eq.\ (\ref{omega3d})
reduces to an expression derived within the TF theory \cite{sedrakian,cozzini}.
This means that Eq.\ (\ref{omega3d}) is valid also in the slow rotation
regime even though this expression is derived assuming the system is in
the mean-field LLL regime. The reason for this validity is explained
in the Appendix of Ref.\ \cite{breathing}.
\bibitem{note_error} The expression of the mode frequency 
[second equation of Eq.\ (A16)] in Ref.\ \cite{breathing} is inadvertently
mistaken.
A factor $(1-Z_0^{-4})$ should be inserted in the last term in the square root.
However, this correction does not affect subsequent discussions 
in that article.
\bibitem{note transverse} There is, of course, a small change in the
transverse breathing mode frequency $\omega_{\rm tr}$ due to the neglected 
$1-(\Omega_0^2/\omega_{\perp}^2)$ term of Eq.\ (\ref{omega3d}).
For $\omega_{\perp}>\omega_z$, $\omega_{\rm tr}$
is given by the upper (positive) sign and thus $\omega_{\rm tr}$
decreases to $2\omega_{\perp}$ with increasing $\Omega_0$.
However, this change is quite small in all the cases considered in this paper.
\bibitem{note_tilt} Here the ratio $\langle z^2 \rangle_{\rm ex,app}^{1/2}/
\langle z^2 \rangle_{\rm ex,true}^{1/2}$ 
between the apparent axial rms radius $\langle z^2 \rangle_{\rm ex,app}^{1/2}$
and the true one $\langle z^2 \rangle_{\rm ex,true}^{1/2}$
is estimated by, assuming an ellipsoidal cloud 
with the aspect ratio of 
$\langle z^2 \rangle_{\rm ex,app}^{1/2}/\langle y^2 \rangle_{\rm ex}^{1/2}$,
the increase of the apparent axial size when the rotation
axis is tilted by $5^\circ$ in the direction of the line of sight.
We mention that, in the fast rotation limit, 
the effect of the axial tilt might be slightly larger than 
the present estimate since the form of the cloud is close to a disk 
rather than an ellipsoid; i.e., the particle density concentrates 
close to the equatorial plane in the Gaussian axial density profile.
\end{thebibliography}
\end{document}